\documentclass[preprints,article,submit,moreauthors]{mdpi} 
\pubvolume{1}
\issuenum{1}
\articlenumber{0}
\pubyear{2026}
\copyrightyear{2026}
\datereceived{ } 
\daterevised{ } 
\dateaccepted{ } 
\datepublished{ } 
\Title{Integration of the NIST Spectroscopic Plate Archive into the Harvard Plate Stacks}
\Author{%
Jacob Ward $^{1,}$$^{2}$*\orcidA{},\:
Thom Burns $^{3}$\orcidD{},\:
Lisa Bravata $^{3}$\orcidE{},\:
Gillian Nave $^{2,}$$^{4}$\orcidB{},\:
and Adam Foster $^{3}$\orcidC{}
}
\AuthorNames{Jacob Ward, Thom Burns, Lisa Bravata,  Gillian Nave, Adam Foster}
\address{%
$^{1}$ \quad Department of Astronomy, University of Maryland College Park, USA\\
$^{2}$ \quad NASA Goddard Space Flight Center, Greenbelt, MD, USA\\
$^{3}$ \quad Center for Astrophysics | Harvard \& Smithsonian, Cambridge, MA, USA\\
$^{4}$ \quad Department of Physics, Imperial College London, UK}
\corres{Correspondence: jwward@umd.edu}
\abstract{%
For many decades, the National Institute of Standards and Technology (NIST) Atomic Spectroscopy Group housed a growing collection of thousands of photographic glass plate negatives. With the closure of the NIST Atomic Spectroscopy Group in March 2025, the irreplaceable collection needed a new home. Staff from NIST were contacted by staff from the Harvard Plate Stacks at the Center for Astrophysics | Harvard \& Smithsonian (CfA). The Harvard Plate Stacks is the largest collection of astronomical glass plates in the world, with over 500,000 glass plate negatives from the 1880s to the 1990s. Given the long-standing history of the Harvard Plate Stacks and the expertise of the staff, it was decided that the NIST archive would move to the CfA. We summarize the contents of this collection regarding the instruments used and the spectra measured, and we present the tools and procedures we use for digitizing the collection. This includes the use of calibrated flatbed scanners for digitizing the plates themselves, as well as the digitization of correlated lab notebooks with important metadata associated with the plates.
}
\begin{document}
\section{Introduction}

For over one hundred years, glass photographic plates have been used to image astronomical objects and their spectra. These plates have been essential for understanding astronomical phenomena and the details of atomic structure alike: a unified tool for linking the smallest and largest features of our universe together. From the 1800s through the 21st century, multiple institutions have amassed collections of these plates from a variety of specialized instruments. One of these collections was housed at the National Institute of Standards and Technology (NIST) Atomic Spectroscopy Group (ASG). With the closure of the ASG in March 2025 and its subsequent move to NASA Goddard Space Flight Center, it was decided that the NIST collection of spectroscopic plates would move to the Harvard Plate Stacks at the Center for Astrophysics | Harvard \& Smithsonian (CfA) \cite{plate_stacks}. The majority of the plates were transferred in April 2025, with a subsequent batch sent in December 2025. 

In January 2026, Jacob Ward and Gillian Nave visited the CfA to begin an inventory of the whole collection. It was clear that it included many plates that may have remaining scientific value, either through their re-measurement and analysis using modern techniques and wavelength standards, or because the plates were only partially analyzed. For example, many plates of cerium spectra were found -- an element of current interest for the interpretation of spectra of kilonovae -- but it appears that linelists for this important element were never published; only the energy levels derived from them were. There are many plates of moderately-ionized spectra that were taken for one ionization stage, but include spectra of adjacent stages that were not published. 

This paper describes what we have found so far in the collection and our future plans for it. We include in section \ref{Zr} an example of how the re-measurement of a plate can result in better wavelengths for the spectra of Zr~I and Zr~II that currently date from the 1930's. 

\section{Collection Contents}

The archive contains measurements of spectra from a large fraction of the periodic table, taken not only at NIST, but at other institutions including the Massachusetts Institute of Technology (MIT), the California Institute of Technology, the University of California, Berkeley, Argonne National Laboratory, and other institutes. A complete inventory of the collection was not undertaken while it was at NIST, and it has only just begun at the Harvard Plate Stacks. The following subsections have been identified primarily from the historical labels on the individual boxes and crates used to store the plates, as well as correspondence and publications.

\subsection{NIST}

The majority of the collection consists of plates from instruments at the National Bureau of Standards (NBS) and NIST. These include the following spectrographs:

\begin{description}
\item{NIST Grazing Incidence (GIVS):} 
The collection holds about 1200 plates from the NIST grazing incidence spectrograph of sliding spark, triggered spark, and laser-produced plasma sources, covering 3 - 70 nm. Many plates were taken for one ionization stage, but may contain spectra of additional ionization stages that remain unpublished.
\item{NIST Normal Incidence Vacuum Spectrograph (NIVS):}
Roughly 2000 plates in the collection were exposed using the NIST Normal Incidence Vacuum Spectrograph (NIVS), which covers the 30-500 nm wavelength range with a reciprocal dispersion of 0.08 nm/mm. Many of these plates have only partially been analyzed. The plates include spectra of neutral through nine-times ionized elements from across the periodic table. One notable set of plates are spectra of Pt/Ne hollow cathode lamps that were used for the calibration of spectrographs on the Hubble Space telescope \cite{pt_atlas, COS_paper}.
\item{NIST Air Eagle:}
There are also roughly 200 plates from the NIST Air Eagle spectrograph, covering 200 - 1000 nm.
\item{Fabry-Per{\'o}t plates:} 
Several boxes contain Fabry-Per{\'o}t spectra of $^{86}$Kr and $^{198}$Hg, taken by V. Kaufmann \cite{Kaufman_1962}. These species were used as primary and secondary wavelength standards for many decades, and thus many of the older spectra in the NIST Atomic Spectra Database (ASD) are likely to have been calibrated with respect to these plates. 
\item{Infrared plates:} Two boxes of infrared plates were found that appear to have been used in a 1927 publication \cite{Kiess_1927}. These are some of the earliest infrared atomic spectra that have been recorded. 
\item{Plates used for `Tables of Spectral Line intensities' \cite{Meggers_1961}:} The intensities of spectral lines from 70 elements in an arc were estimated from these plates and some are still used in the Atomic Spectra Database, particularly where modern compilations of atomic data have not been made. 
\end{description}

\subsection{MIT}

Beginning in roughly 1940, MIT constructed a 100,000 Gauss electromagnet for studying the Zeeman effect on spectra of a variety of ions. The source was a specially-constructed horizontal arc that produced spectra of neutral through doubly-ionized atoms. On account of the large power consumption of the magnet, it was desired to shorten runs as much as possible, and arrangements were made to send light from the one arc simultaneously into three 35-foot concave diffraction-grating spectrographs. Each run thus produced between 12 and 21 plates, with a UV reciprocal dispersion of 0.08 nm/mm.

The collection includes roughly 1000 Zeeman plates that were taken by G. Harrison at MIT and measured at NBS by W.F. Meggers \citep{Harrison_1940}. The plates include UV spectra 
of heavier elements of strong interest for kilonova spectroscopy, including many lanthanide elements. The wavelengths of many of these ions could be improved by re-analysis of the plates using modern measurement techniques and wavelength standards, discussed below in section \ref{Zr}.

\subsection{Spectra from other institutions}
\begin{itemize}
\item About 200 plates of promethium, taken at U. Berkeley by J. Reader in connection with his PhD dissertation, were identified \cite{Reader_1963,Reader_1967}. The high radioactivity of promethium means that these spectra are unlikely to be taken again. Improvements in measuring the plates with modern wavelength standards may give better wavelengths and energy level values.
\item Many boxes contain cerium spectra, with a variety of ionization stages and from a variety of instruments at various institutions in the USA \cite{Verges_1972,Corliss_1973}. This element is of high interest for the analysis of kilonova spectra. 
\end{itemize}

\subsection{Lab books}

Several dozen laboratory books were found during the course of clearing out the rooms at NIST. These include books connected with measurements using the NIST GIVS and NIVS spectrographs, all of which are clearly labeled with dates and serial numbers, so it should be possible to link these to plates in the collection. There are many other books from NBS/NIST which we have yet to link to either instruments, researchers, or plates. Digitizing these books will be an important task in completing the inventory of the collection and can be completed at Harvard and made freely available online.

\section{Plans for the collection}

Our first priority is to complete a high level inventory of the collection which is currently about halfway complete. The next priority is to digitize the lab books and then rehouse the plates in archival sleeves and cabinets. During this process, we gain a reasonably good idea of the contents of the archive and can proceed to make these publicly available. While our long-term goal is to digitize the entire collection of plates, we do currently do not have the resources or dedicated staffing to complete this. Digitizing a few plates requested by the community would be possible, but collaborations with the outside community or external grant funding will be essential for a more complete digital archive. 

\subsection{Condition of the plates}

The plates are in a variety of conditions, depending on their age and storage conditions. They are currently housed in a variety of sleeves and boxes that are often the original manufacturer's boxes for the negatives, or even wooden crates, none of which are satisfactory for long-term storage of photographic materials. It is important to house all of the plates in buffered acid-free paper sleeves that have been Photographic Activity Tested (PAT) in subdivided archival boxes, organized in powder coated metal cabinets so that the plates will be preserved for many decades.

\begin{figure}[H]
\isPreprints{\centering}{}
\subfloat[\centering]{\includegraphics[width=12.0 cm]{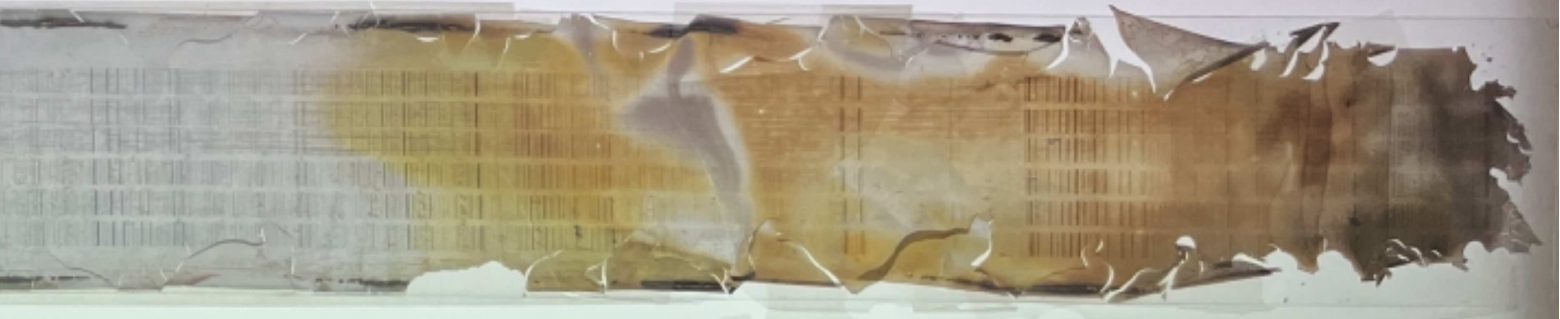}}\\
\subfloat[\centering]{\includegraphics[height=12.0 cm,angle=270]{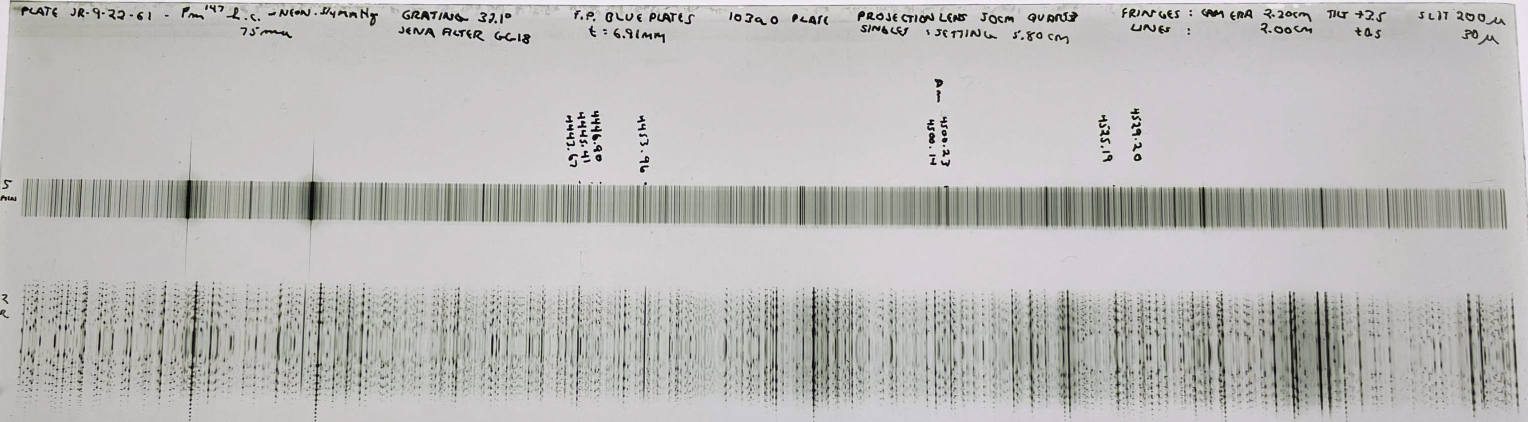}}
\caption{(a) Example of a plate that has degraded after being stored in a plastic sleeve. (b) Example of a plate that has not been significantly degraded.\label{fig:badplate}}
\end{figure}

Figure \ref{fig:badplate}a shows the effects of poor storage. This plate is stored in a plastic sleeve common from the 1960s-2000s. The plastic initially protected it from dust and abrasion while still allowing the spectra and writing to be viewed without removal. Within as little as three decades, the plastic has deteriorated in a process known as vinegar syndrome. This is an off-gassing reaction detectable by the plastic becoming brittle and the smell of acetic acid. Because the plastic remains largely impermeable, even after deterioration, it traps these gasses and humidity, creating a microclimate. The acetic acid reacts with the silver-gelatin photographic emulsion, causing irreversible yellowing and browning. The humid microclimate, when combined with moderate temperature fluctuations, causes the gelatin emulsion to delaminate from the glass, resulting in flaking \cite{flemming_plates}. This is why plates from the 1900-1950s stored in acidic paper enclosures can show less damage than plates from the 1980s stored in plastic sleeves. The off-gassing also accelerates deterioration of other nearby plastics. For this reason, the plates should be stored in open shelving without lids until they can be properly rehoused \cite{harvard_library_glass_plates}.

To recover some information from the plate shown in \ref{fig:badplate}a, it needs to be imaged before it is removed from the plastic sleeve and directly after, so both written metadata and flaking emulsion can be captured, though many spectral lines are no longer measurable. The lower plate, dating from 1961,  contains grating and Fabry-Per{\'o}t spectra of promethium and is generally in good condition, even though it is stored without any sleeve and stacked on top of other plates in an old paper box. Re-measuring and re-analyzing this plate is much more viable.

\subsection{Digitization}\label{digit}

The Harvard Plate Stacks has two imaging systems that can be used to digitize the entire plate collection. 
The first is a commercial scanner (Epson XL11000) with an optical resolution of 10.6~$\mu$m (2400 pixels/in), and a scanner bed measuring 420~mm by 310~mm. This was used to evaluate the periodic errors and resolution of commercial scanners for spectroscopic plates \cite{Wyatt_2017}, showing that large periodic errors can be eliminated by placing the plate with the dispersion direction parallel to the short side (310~mm) side of the scanner. The measured resolution and position accuracy is comparable with rotating prism comparators previously used to scan photographic plates at NIST. 

The second is a multi-spectral imaging (MSI) system that digitizes plates across multiple wavelength bands. The system consists of a 150-megapixel iXH Achromatic PhaseOne Camera with a Schneider 72mm Mark I lens coupled with transmissive and reflective arrays of 16 specific narrow-wavelength light-emitting diodes (LEDs) in the UV, visible, and IR wavelengths. This system was designed for direct imaging of astronomical glass plates and has yet to be evaluated to assess periodic errors across a 2 x 18-inch spectral plate \cite{advanced_plates}. MSI can be used to distinguish features in the silver halide crystals of the photographic emulsion from other obscuring features, such as deterioration effects in the emulsion and handwritten ink annotations \cite{ink_plates}. It has also been shown to be useful in identifying characteristics of the glass and emulsion that might help identify age and condition issues.

\section{Potential Scientific value: A re-analysis of zirconium spectra}\label{Zr}

\begin{figure}[H]
\isPreprints{\centering}{}
\includegraphics[height=13.5 cm,angle=270]{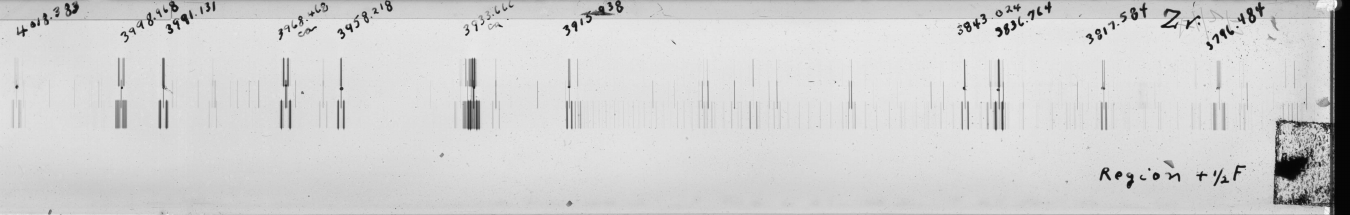}
\caption{Example of a plate from the MIT collection of Zeeman spectra of zirconium. The central, zero field track was used to derive improved wavelengths for Zr~I and Zr~II.\label{fig_plate}}
\end{figure}

A number of boxes of plates from the MIT collection were left behind at NIST after the initial shipment, and one plate of Zeeman spectra of zirconium from these boxes was selected for analysis, to see if improved wavelengths could be derived by re-scanning the plate using the flatbed scanner and calibrating it using modern wavelength standards (See Fig. \ref{fig_plate}). We have been unable to match this plate to any publication of zirconium spectra, but the description of the experiment is given in \cite{Harrison_1940}.  The plate has three tracks, the central of which is the spectrum of an arc recorded without a magnetic field, and the outer two are the $\pi$ and $\sigma$ components. The central, zero field track has a reciprocal dispersion of 0.08~nm/mm, covers the wavelength range 379 to 402~nm, and was the track measured for the re-analysis. It was digitized using the flatbed scanner mentioned in section \ref{digit} and the positions of the lines were measured using the ASPAS program \cite{ASPAS}.

The wavelengths for both Zr~I and Zr~II in ASD are taken from the MIT wavelength tables \cite{Harrison_1969} and have uncertainties of at least 0.005~nm. The most recent publication on wavelengths and energy levels of Zr~I and Zr~II is that of Lawler et al. \cite{Lawler_2022}, which gives improved wavelengths for 372 lines of Zr~I and 89 lines of Zr~II, based on Fourier transform spectroscopy of hollow cathode lamps. The measured Zr~I wavelengths of \cite{Lawler_2022} have an uncertainty of roughly 4$\times10^{-5}$~nm in the wavelength region covered by the MIT plate, over an order of magnitude lower than the estimated uncertainty of 5$\times10^{-4}$~nm quoted in \cite{Harrison_1969}. No explicit uncertainties are given in the wavelengths from ASD, but are estimated as between 2.5 and 25 in the last digit, or between 2.5$\times10^{-3}$ and 2.5$\times10^{-2}$~nm. 

\begin{figure}[H]
\isPreprints{\centering}{}
\includegraphics[height=4.5 cm]{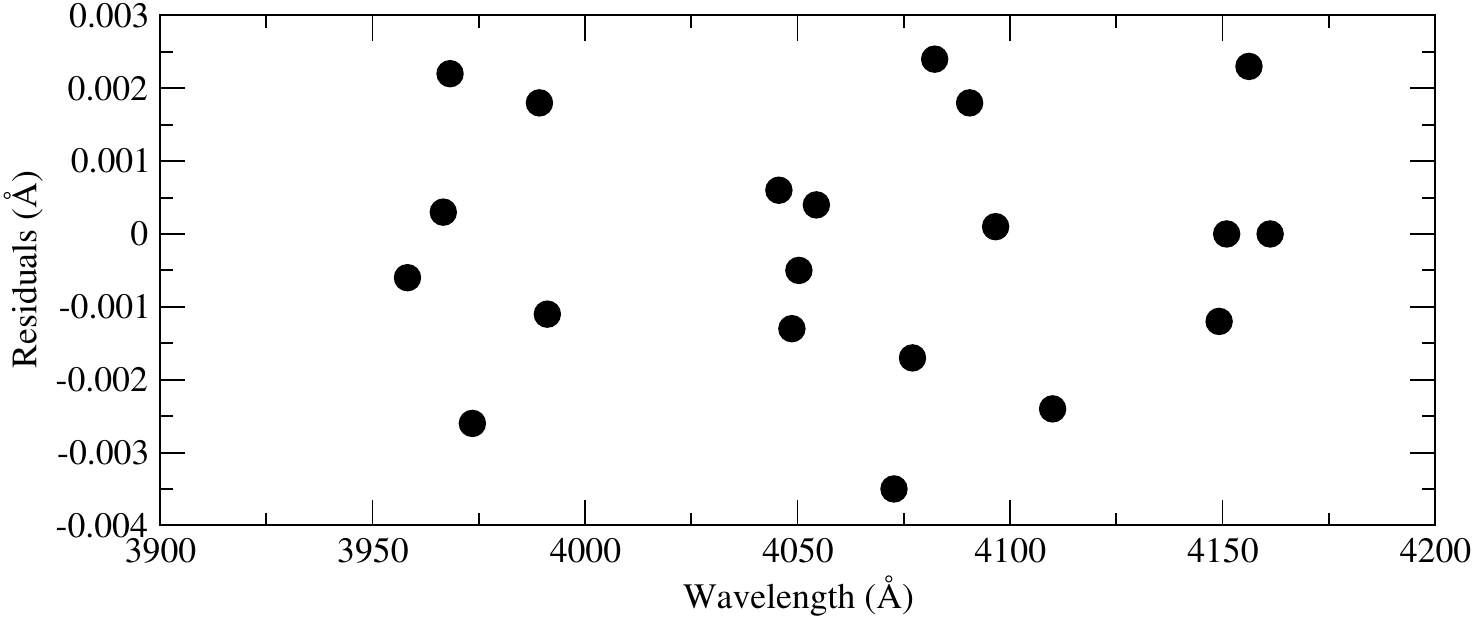}
\caption{Residuals from a fifth-order fit of 19 wavelengths measured on a zirconium plate from the MIT collection to those of \cite{Lawler_2022}.\label{fig_resid}}
\end{figure}

The tables in \cite{Lawler_2022} include 19 of the 97 lines we measured on the plate. We used these 19 lines to fit a 5th order polynomial to the line positions and the residuals from this fit are given in figure \ref{fig_resid}. The standard deviation of these residuals is 1.6$\times 10^{-4}$~nm, and we take this as an estimate of the statistical uncertainty of the wavelengths. This is roughly a factor of four better than the estimated uncertainty from \cite{Harrison_1969}, and we attribute the improvement to better measurements of the centroid of the spectral lines and the improved wavelength references from \cite{Lawler_2022}. Using these re-measured lines, we can then derive improved wavelengths for the remaining 78 lines that are not present in \cite{Lawler_2022}. 

\section{Conclusions}

The contents of the archive of spectroscopic plates formerly held by NIST have now been transferred to the Harvard Plate Stacks at the CfA. Many plates contain spectra that are still of scientific value, either because they have been only partially analyzed, or because modern techniques and wavelength standards can be used to improve the uncertainty in the wavelengths by an order of magnitude or more. While the partial catalog of the collection reveals some notable plates, such as those of high interest for kilonova spectroscopy, the scientific value of the whole set is difficult to summarize. A single plate chosen at random, as described above in section \ref{Zr}, yielded notable improvements to current atomic data. Scaled to the entirety of the multiple-thousands plate collection, the opportunity for progress is immense given proper support and community involvement.

\vspace{6pt}

\authorcontributions{Writing--original draft: Gillian Nave and Jacob Ward; writing--review and editing: Thom Burns, Adam Foster and Lisa Bravata.} 


\funding{This research received no external funding.}

\dataavailability{Data sharing is not applicable.}

\acknowledgments{We acknowledge Albert Henins, Rodrigo Ibacache, and Aung Naing for their help with moving the collection from NIST to CfA. }

\conflictsofinterest{The authors declare no conflicts of interest.} 

\reftitle{References}
\bibliography{refs.bib}

@misc{plate_stacks,
	title={The Harvard Plate Stacks},
	note={Accessed: 2026-09-01},
	howpublished={\url{https://platestacks.cfa.harvard.edu}}
}

@misc{ASPAS,
  title={ASPAS},
  note={Accessed: 2026-09-01},
  howpublished = {\url{https://github.com/gnave/ASPAS}}
}

@misc{harvard_library_glass_plates,
  title={Preservation Recommendations for Historic Glass Astronomical Plates},
  author={{Harvard Library Preservation Services}},
  note={Accessed: 2026-09-14},
  howpublished = {\url{https://preservation.library.harvard.edu/historic-glass-astronomical-plates}}
}

@article{flemming_plates,
  title={A Legacy of Stars: Preservation of the Williamina Fleming Astronomical Glass Plates from Harvard College Observatory},
  author={Elena Bulat and Tess Bronwyn Hamilton and Thom Burns and Samara Ayvazian-Hancock and Debora Mayer and Arthur McClell and Georgina Rayner and Amanda Maloney},
  volume={21},
  journal={Topics in Photographic Preservation (forthcoming)},
  publisher={American Institute for Conservation}
}

@inproceedings{advanced_plates,
  author = {Toth, Michael B. and Lisa Hees and Thom Burns and Debora D. Mayer and Kayleigh MacDonald},
  title = {Integrating Advanced Imaging of Astronomical Glass Plates},
  booktitle = {In Proceedings of the 54th American Institute for Conservation (AIC) Annual Meeting},
  year = {2026}
}

@article{ink_plates,
  title={Technical Analysis of Inks Used for Scientific Annotations on the Harvard Astronomical Photographic Glass Plate Collection},
  author={Elena Bulat and Thom Burns and Tess Bronwyn Hamilton and Debora Mayer and Arthur McClelland and Georgina Rayner and Samara Ayvazian-Hancock and Amanda Maloney},
  volume={21},
  journal={Topics in Photographic Preservation (forthcoming)},
  publisher={American Institute for Conservation}
}

@article{pt_atlas,
  journal={J. Res. Natl. Inst. Stand. Technol.},
  volume={97},
  pages={1--211},
  title={Atlas of the Spectrum of a Platinum/Neon Hollow-Cathode Reference Lamp in the Region 1130-4330 \AA},
  author={J. E. Sansonetti and J. Reader and C. J. Sansonetti and N. Acquista},
  year={1992},
  doi={10.6028/jres.097.002}
}

@article{COS_paper,
doi = {10.1086/668813},
url = {https://doi.org/10.1086/668813},
year = {2012},
publisher = {University of Chicago Press},
volume = {124},
number = {922},
pages = {1295},
author = {Nave, Gillian and Sansonetti, Craig J. and Penton, Steven V. and Cunningham, Nathaniel and Beasley, Matthew and Osterman, Steven and Kerber, Florian and Keyes, Charles D. (Tony) and Rosa, Michael R.},
title = {Lifetime and Failure Characteristics of Pt/Ne Hollow Cathode Lamps Used as Calibration Sources for UV Space Instruments},
journal = {Publications of the Astronomical Society of the Pacific}
}

@ARTICLE{Kaufman_1962,
       author = {{Kaufman}, Victor},
        title = "{Wavelengths, Energy Levels, and Pressure Shifts in Mercury 198}",
      journal = {Journal of the Optical Society of America (1917-1983)},
         year = 1962,
       volume = {52},
       number = {8},
        pages = {866},
          doi = {10.1364/JOSA.52.000866},
       adsurl = {https://ui.adsabs.harvard.edu/abs/1962JOSA...52..866K}
}

@article{Kiess_1927,
  journal={Bur. Stand. J. Res.},
  volume={1},
  pages={75--90},
  title={Interferometer Measurements of Wave Lengths in the Vacuum Arc Spectra of Titanium and Other Elements},
  author={C. C. Kiess},
  year={1928},
  doi={10.6028/jres.001.004}}

@misc{Meggers_1961,
  author = {William F Meggers and Charles H Corliss and Bourdon F Scribner},
  title = {Tables of spectral-line intensities ::part 1},
  year = {1961},
  publisher = {National Institute of Standards and Technology, Gaithersburg, MD},
  doi = {https://doi.org/10.6028/NBS.MONO.32-1}
}

@ARTICLE{Harrison_1940,
       author = {{Harrison}, George R. and {Bitter}, Francis},
        title = "{Zeeman Effects in Complex Spectra at Fields up to 100,000 Gauss}",
      journal = {Physical Review},
         year = 1940,
       volume = {57},
       number = {1},
        pages = {15-20},
          doi = {10.1103/PhysRev.57.15},
       adsurl = {https://ui.adsabs.harvard.edu/abs/1940PhRv...57...15H}
}

@ARTICLE{Reader_1963,
       author = {{Reader}, Joseph and {Davis}, Sumner P.},
        title = "{Promethium 147 Hyperfine Structure under High Resolution*}",
      journal = {Journal of the Optical Society of America (1917-1983)},
         year = 1963,
       volume = {53},
       number = {4},
        pages = {431},
          doi = {10.1364/JOSA.53.000431},
       adsurl = {https://ui.adsabs.harvard.edu/abs/1963JOSA...53..431R}
}

@article{Reader_1967,
  journal={J. Res. Natl. Bur. Stand. (U.S.), Sect. A},
  volume={71},
  pages={587--599},
  title={Fundamental Energy Levels of Neutral Promethium (Pm~I)},
  author={J. Reader and S. Davis},
  year={1967},
  doi={10.6028/jres.071A.050}}

@article{Verges_1972,
  journal={J. Res. Natl. Bur. Stand. (U.S.), Sect. A},
  volume={76},
  pages={285--304},
  title={Infrared Spectra of Cerium (Ce~I and Ce~II) Between 0.8 and 2.4~$\mu$m},
  author={J. Verges and C. H. Corliss and W. C. Martin},
  year={1972},
  doi={10.6028/jres.076A.030}}

@article{Corliss_1973,
  journal={J. Res. Natl. Bur. Stand. (U.S.), Sect. A},
  volume={77},
  pages={419--546},
  title={Wavelengths and Energy Levels of the Second Spectrum of Cerium (Ce~II)},
  author={C. H. Corliss},
  year={1973},
  doi={10.6028/jres.077A.032},
  notes={NIST compilation}}

@article{Wyatt_2017,
  journal={Appl. Opt.},
  volume={56},
  pages={3744--3749},
  title={Evaluation of Resolution and Periodic Errors of a Flatbed Scanner Used for Digitizing Spectroscopic Photographic Plates},
  author={M. Wyatt and G. Nave},
  year={2017},
  doi={10.1364/AO.56.003744},
  notes={instrumentation},
  keywords_el={}}

@book{Harrison_1969,
  booktitle={M.I.T. Wavelength Tables},
  edition={1969},
  publisher={MIT Press},
  address={Cambridge, Massachusetts},
  title={M.I.T. Wavelength Tables},
  author={G. R. Harrison},
  year={1969}}

@article{Lawler_2022,
  journal={J. Quant. Spectrosc. Radiat. Transfer},
  volume={289},
  pages={108283},
  title={Energy Levels of Singly Ionized and Neutral Zirconium},
  author={J. E. Lawler and J. R. Schmidt and E. A. Den Hartog},
  year={2022},
  doi={10.1016/j.jqsrt.2022.108283},
  notes={fnl_2022}}

\PublishersNote{}
\end{document}